\documentclass[10pt,twocolumn]{article}

\usepackage{natbib}

\usepackage{graphicx}

\usepackage{txfonts}

\usepackage{amssymb}
\usepackage{geometry}
\begin{document}

\title{On the emission polarization of RX J1856.5-3754}
\author{Nino Chkheidze\thanks{E-mail:
n.chkheidze@gmail.com}\\ Tbilisi State University, Chavchavadze
Avenue 3, 0128, Tbilisi, Georgia}

\date{}

\maketitle


\begin{abstract}The polarization properties of RX J1856.5-3754 are
investigated, based on the plasma emission model presented in
previous works. It is shown that if the emission of this source is
generated by the synchrotron mechanism, the X-ray and the optical
emission are linearly polarized and the corresponding degrees of
linear polarization are calculated. The measurement of the
polarization patterns of this source appears to be most reliable
way, to make clear its real emission nature.
\end{abstract}

%

\section{Introduction}

RX J1856.5-3754 (or RXJ1856, hereafter) is one of the brightest,
nearby isolated neutron stars discovered by ROSAT \citep{walt96} as
an X-ray source and a large number of papers have been devoted to
explaining its nature. Its soft X-ray spectrum can be well fit by
the Planckian spectrum with a temperature
\(kT_{bb}^{\infty}\simeq63\pm3\)eV \citep{burw03}. Thus it is
supposed that the emission of this source is of thermal nature.
However, the existing models based on this original assumption face
problems.

The lack of any significant spectral lines in the X-ray spectrum of
RXJ1856 argues against heavy element atmosphere models
\citep{burw01,burw03}, whereas single temperature hydrogen
atmosphere fits over-predict the optical flux by a large factor
\citep{pavlov96,pons02,burw03}. As none of the classic models of
neutron star atmosphere are able to explain the observed X-ray
spectrum, which is well fitted by a simple blackbody model
\citep{burw03}, it has been proposed that the star has no atmosphere
but a condensed matter surface \citep{burw01,turolla04}. This
surface might emit a virtually featureless blackbody-like spectrum
(as originally proposed by \citet{pavlov00}). Since the broadband
spectra of this source can not be fitted by a single Planckian
spectrum, it is often described by two-temperature blackbody models
\citep{pons02,pavlov02,burw03}. However, condensation of surface
matter requires very specific conditions to be fulfilled
\citep{lai97,lai01}. But even if these conditions are satisfied, the
formation of a non-uniform distribution of the surface temperature
(two-temperature blackbody models) still remains unclear. The most
adequate fits of the spectra give models which assume that the star
has a thin hydrogen atmosphere superposed on a condensed matter
surface \citep{mot03,ho07}. However, the origin of such thin
hydrogen layers fitting the data is a problematic issue.

Alternatively, the observational properties of RXJ1856 can be
explained in the framework of the plasma emission model presented in
previous papers \citep{ch07,ch08}, which are based on well-developed
theory of pulsars. This model suggests that the emission from this
object is generated by synchrotron radiation, created as the result
of acquirement of pitch angles by relativistic electrons during the
quasi-linear stage of the cyclotron instability. The model gives
successful fits for broadband spectra, without facing the problems
typical for the thermal radiation models. Considering the case of a
nearly aligned rotator, it was predicted that the source should have
pulsated with a period of $\sim1$ s \citep{ch07}. However, the
posterior XMM-Newton observation of RXJ1856 discovered that its
X-ray emission pulsates with a period of $7.055$ s \citep{tien07}.
The latter fact has been explained in the framework of drift wave
driven model \citep{ch08}. In particular, the real spin period of
the pulsar might differ from the observable one, as a consequence of
the existence of very low frequency drift waves in the region of
generation of the pulsar emission. These waves are not directly
observable but only result in a periodical change of curvature of
the magnetic field lines and, hence, a periodical change of the
emission direction with a period of the drift waves assumed to be
equal to the observable period \citep{lom06}.

We are not about to reject the existing thermal radiation models for
RXJ1856 and suppose that the most reliable argument for revealing
the real emission nature of this source will be its study with
polarization instruments. Therefore in present paper the
polarization properties of RXJ1856 are investigated, in the
framework of the plasma emission model developed in previous works
\citep{ch07,ch08}.

In this paper, we give a description of the emission model in (Sect.
2), investigate the emission polarization in (Sect. 3) and discuss
our results in (Sect. 4).

\section{Emission mechanism}

It is well known that the distribution function of relativistic
particles is one dimensional at the pulsar surface, because any
transverse momenta (\(p_{\perp}\)) of relativistic electrons are
lost in a very short time (\(\leq10^{-20}\)s) via synchrotron
emission in very strong \(B\sim10^{11}\)G magnetic fields. For
typical pulsars, the plasma consists of the following components:
the bulk of plasma with an average Lorentz-factor
$\gamma_{p}\simeq10$; the tail of the distribution function with
$\gamma_{t}\simeq10^{4}$ and the primary beam with
$\gamma_{b}\simeq10^{6}$ (see Fig.1 from \citet{arons81}).
Generation of waves is possible during the further motion of the
relativistic particles along the dipolar magnetic field lines if the
condition of cyclotron resonance is fulfilled \citep{kaz91b}:
\begin{equation}
    \omega-k_{\varphi}V_{\varphi}-k_{x}u_{x}+\omega_{B}/\gamma_{r}=0,
    \end{equation}
where \(k_{\varphi}^{2}+k_{\perp}^{2}=k^{2}\),
\(k_{\perp}^{2}=k_{x}^{2}+k_{r}^{2}\), $V_{\varphi}$ is the particle
velocity along the magnetic field, $\gamma_{r}$ is the
Lorentz-factor for the resonant particles,
$u_{x}=cV_{\varphi}\gamma_{r}/\rho\omega_{B}$ is the drift velocity
of the particles due to curvature of the field lines, $\rho$ is the
radius of curvature of the field lines and $\omega_{B}=eB/mc$ is the
cyclotron frequency. Here a cylindrical coordinate system is chosen,
with the $x$-axis directed perpendicular to the plane of field line,
when $r$ and $\varphi$ are the radial and azimuthal coordinates. The
reason for the generation of such waves is an anisotropy of the
distribution function. During the quasi-linear stage of the
instability, a diffusion of particles arises not only along but also
across the magnetic field lines. The resonant electrons, therefore,
acquire transverse momenta and, as a result, start to radiate
through the synchrotron mechanism.

In \citet{ch07}, it has been assumed that the emission of RXJ1856 is
generated by the synchrotron mechanism, which switches on during the
cyclotron instability developed at distances $r\sim10^{9}$cm. The
observed X-ray spectrum is a result of radiation of the primary-beam
electrons. Particularly, the synchrotron spectrum of the beam
electrons with the energy distribution:
\begin{equation}\label{11}
    \textit{f} _{\parallel b}=\frac{n_{b}}{\sqrt{\pi}\gamma_{T}}\textrm{ exp}\left[-\frac{\left(\gamma_{r}-\gamma_{b}\right)^{2}}{\gamma_{T}^{2}}\right],
\end{equation}
well matches the measured X-ray spectrum  in the energy interval
$(0.26-0.9)$keV (here $\gamma_{T}\simeq10$ - is the half-width of
the distribution function, $n_{b}=B/Pce$ is the density of primary
beam electrons, equal to the Goldreich-Julian density \citep{gold69}
and $P$ is the pulsar spin period. The wave excitation leads to a
redistribution of the resonant particles via quasi-linear diffusion.
Consequently, by achieving the stationary mode the distribution
function of the beam electrons takes the following form:
\begin{equation}
    \textit{f}_{\parallel b}\propto\gamma_{r}^{-4}.
\end{equation}
At the same time the radiation density appears to be sufficiently
high and prevents the domination of the self-absorption processes.
Synchrotron self-absorption redistributes the emission spectrum and,
in the energy domain of relatively low frequencies it experiences
drop. Hence, for the energy interval $(0.15-0.26)$keV the X-ray
emission spectrum  has the form $\nu^{5/2}$, which fits well the
observed one (see Fig.4 from \citet{ch07}). The measured optical
spectrum is the result of the synchrotron emission of the tail
electrons, redistributed due to cyclotron instability:
\begin{equation}
    \textit{f}_{\parallel t}\propto\gamma_{r}^{2}.
\end{equation}
The resulting theoretical spectrum matches the measured one closely.

\section{Emission polarization}

\begin{figure}
\centering
 \includegraphics[width=2.5 cm]{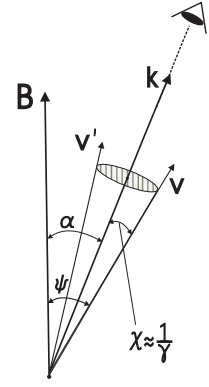}
 \caption{{\bf $\textbf{k}$}
is the emission direction, {\bf \textbf{B}} - the magnetic field
line,  \textbf{v} and {\bf $\textbf{v}^\prime$} - are the electrons'
velocities.}
       \label{Fig1}
\end{figure}

Now let us consider the polarization properties of RXJ1856. For this
reason, we have to find the Stokes parameters, which, in our case,
are defined as follows \citep{ginz81}:
\begin{eqnarray}
    I(\nu)=\frac{3e^{3}}{2\pi}\int B\left(\frac{\nu}{\nu_{c}}\right)^{2}(1+\chi^{2}\gamma^{2})\gamma\textit{f}(\gamma,\psi)\sin\alpha\times\nonumber
    \\\times\left[(1+\chi^{2}\gamma^{2})K_{2/3}^{2}(g_{\nu})+\chi^{2}\gamma^{2}K_{1/3}^{2}(g_{\nu})\right]d\chi
d\gamma,
    \label{Eq3}
\end{eqnarray}
\begin{eqnarray}
   Q(\nu)=\frac{3e^{3}}{2\pi}\int
   B\left(\frac{\nu}{\nu_{c}}\right)^{2}(1+\chi^{2}\gamma^{2})\gamma^{2}\textit{f}(\gamma,\psi)\sin\alpha\cos2\tilde{\chi}\times\nonumber\\
   \times\left[(1+\chi^{2}\gamma^{2})K_{2/3}^{2}(g_{\nu})-\chi^{2}\gamma^{2}K_{1/3}^{2}(g_{\nu})\right]d\chi d \gamma,
\end{eqnarray}
\begin{eqnarray}
   U(\nu)=\frac{3e^{3}}{2\pi}\int
   B\left(\frac{\nu}{\nu_{c}}\right)^{2}(1+\chi^{2}\gamma^{2})\gamma^{2}\textit{f}(\gamma,\psi)\sin\alpha\sin2\tilde{\chi}\times\nonumber\\
   \times\left[(1+\chi^{2}\gamma^{2})K_{2/3}^{2}(g_{\nu})-\chi^{2}\gamma^{2}K_{1/3}^{2}(g_{\nu})\right]d\chi d \gamma,
\end{eqnarray}
\begin{eqnarray}
   V(\nu)=\frac{3e^{3}}{\pi}\int
   B\left(\frac{\nu}{\nu_{c}}\right)^{2}(1+\chi^{2}\gamma^{2})^{3/2}\gamma^{2}\chi\textit{f}(\gamma,\psi)\sin\alpha \times\nonumber\\
   \times K_{1/3}(g_{\nu})  K_{2/3}(g_{\nu})d\chi d \gamma,
\end{eqnarray}
where
\begin{equation}
   \nu_{c}\approx4\cdot10^{3}B\gamma^{2}, \qquad \qquad
    g_{\nu}=\frac{\nu}{2\nu_{c}}(1+\chi^{2}\gamma^{2})^{3/2},
\end{equation}

\begin{equation}
   \textit{f}(\gamma,\psi)=\textit{f}_{\parallel}(\gamma)\textit{f}_{\perp}(\psi).
\end{equation}
Here $\alpha$ is the angle between $ \textbf{B}$ and $ \textbf{k}$,
$\psi$ - is the pitch angle, $\chi$ is the angle between the wave
vector and the electron velocity vector (see Fig.~\ref{Fig1}). Also,
$\gamma$ is the Lorentz-factor of emitting electrons,
$\textit{f}_{\parallel}(\gamma)$ and $\textit{f}_{\perp}(\psi)$ are
the distribution functions of the emitting particles by parallel and
perpendicular momenta, respectively, $K_{2/3}(g_{\nu})$ and
$K_{1/3}(g_{\nu})$ are the Macdonald functions and angle
$\tilde{\chi}$ defines the direction of maximum intensity of the
polarized component on the plane of sky. According to \citet{ch07}
the following condition is fulfilled for the resonant particles:
$\psi_{0}\gamma_{r}\gg1$ (where $\psi_{0}$ is the mean value of the
pitch angle). In this case, the emitting particles' distribution
function by their pitch angles has the following form
\citep{malov02}:
\begin{equation}
    \qquad \textit{f}_{\perp}(\psi)\propto e^{-A\psi^{4}},\qquad
    A=\frac{4e^{6}}{3\pi^{3}m^{5}c^{7}}\frac{B^{4}P^{3}\gamma_{p}^{4}\gamma_{r}^{2}}{\gamma_{b}^{3}}.
   \end{equation}

\begin{figure}
\centering
\includegraphics[width=4 cm]{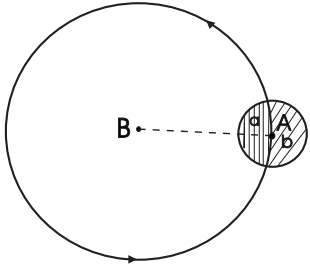}
\caption{Point {\bf $\textbf{B}$} corresponds to the magnetic field
line and  point {\bf \textbf{A}} - to the observer's line of sight.
 }
       \label{Fig2}
\end{figure}

Synchrotron emission of a single electron is strongly beamed along
the direction of motion into an angle of order $1/\gamma$. Hence, in
the given direction, the observer will detect radiation of the
electrons with velocities filling up the cone with $1/\gamma$ in
angular size and with the major axis coincided with the line of
sight of an observer (see Fig.~\ref{Fig1}). Therefore, at the given
moment of time, the observer receives emission of the electrons
having the pitch-angles from the interval
$\psi=\alpha+\chi\subset[\alpha-1/\gamma, \alpha+1/\gamma]$ (it
should be mentioned that the angle $\chi$ is considered to be
positive if $\psi>\alpha$ and to be negative, otherwise). The Stokes
parameter $V$ is an odd function of variable $\chi$, therefore, its
integration over symmetric interval yields zero. Let us consider the
top view of Fig.~\ref{Fig1} (see Fig.~\ref{Fig2}). The larger circle
represents a cross section of the cone. The lateral surface of this
cone is described by the electron velocity vector moving on a spiral
along the magnetic field line. The circle with smaller radius
corresponds to the cone combined by the velocity vectors giving the
significant radiation in the observer's direction. Its angular size
estimates as $1/\gamma$. The area of the smaller circle is striped
in two different ways: area $\textbf{a}$ corresponds to the
electrons which have positive values of angle $\chi$, while area
$\textbf{b}$ corresponds to those having negative values of angle
$\chi$. As these areas are not equal to each other, one would expect
the nonzero value of $V$. But bearing in mind that the bigger circle
is large enough in comparison with the smaller one (the solid angles
of corresponding cones are of the order of $10^{-3}$ and $10^{-6}$
for the beam electrons and, $10^{-1}$ and $10^{-4}$ for the tail
electrons, respectively), one can easily assume that
$\textbf{a}\approx\textbf{b}$. But only symmetry of integration
bounds is not enough condition to have a zero circular polarization.
Also the distribution function by pitch angles, containing the
variable $\chi$, should change very slowly within the small interval
$\sim1/\gamma$. The estimations show that the latter condition is
well fulfilled. Thus, we conclude that the emission of this source
is not circularly polarized.

Now we can calculate the degree for the linear polarization, which
writes as \citep{ginz81}:
\begin{equation}
   \Pi_{\emph{l}}=\frac{\sqrt{Q^{2}+U^{2}}}{I}.
   \end{equation}
The degree of linear polarization, corresponding to the energy
interval ($0.26 - 0.9$)keV of the observed X-ray spectrum can be
calculated using Eqs. (2), (5), (6), (7) and (11). We get
$\Pi_{\emph{l}}\approx(84-79)\%$ for the energies from the given
interval. As it was mentioned above, the measured X-ray spectrum at
relatively low frequencies $(0.15-0.26)$keV is produced due to
synchrotron self-absorption process, of the beam electrons with the
distribution $\textit{f}_{\parallel b}\propto\gamma_{r}^{-4}$. The
degree of linear polarization in this case is defined as follows:
\begin{equation}
    \Pi_{\emph{l}}=\frac{3}{6d+13},
\end{equation}
where $d=4$. The calculations give $\Pi_{\emph{l}}\approx8\%$. So,
in the energy interval ($0.15 - 0.26$keV), the degree of linear
polarization of the X-ray emission undergoes the strong drop.

The observed optical spectrum is the result of the synchrotron
emission of tail electrons, redistributed due to quasi-linear
diffusion:
\begin{equation}
    \textit{f}_{\parallel_{t}}\propto\gamma^{2}.
\end{equation}
The degree of polarization of the optical emission yields
$\Pi_{\emph{l}}\approx86\%$, in this case.

\section{Discussion}
In previous works \citep{ch07,ch08} the plasma emission model of
RXJ1856 is presented. It is supposed that the emission of this
source is generated by the synchrotron mechanism. The distribution
function of relativistic particles is one dimensional at the pulsar
surface, but plasma with anisotropic distribution function is
unstable that can lead to a wave excitation. The main mechanism of
wave generation in plasmas of the pulsar magnetosphere is the
cyclotron instability. During the quasi-linear stage of the
instability, a diffusion of particles arises as along, also across
the magnetic field lines. Therefore, plasma particles acquire
transverse momenta and, as a result, the synchrotron mechanism is
switched on. The measured X-ray and optical spectra are the results
of the synchrotron emission of primary-beam and tail electrons,
respectively. The predictable characteristic frequencies
\(\nu_{m}(X-ray)\simeq6\cdot10^{16}\)Hz and
\(\nu_{m}(Optic)\simeq2\cdot10^{14}\)Hz (where $\nu_{m}$ is the
frequency of maximum in synchrotron emission spectrum) enter the
same domains as the measured spectra.

The original waves, excited during the cyclotron resonance, come in
the radio domain, but the radio emission is not observed from
RXJ1856. One of the possible explanations why the radio emission is
not detected from this object is that it traverses a large distance
in the pulsar magnetosphere (since the model of a nearly aligned
rotator is used). So there is a high probability for the excited
waves to come in the cyclotron damping range:
$\omega-k_{\varphi}V_{\varphi}-k_{x}u_{x}-\omega_{B}/\gamma_{r}=0$
\citep{khe97}. In this case, the radio emission will not reach an
observer. Nevertheless, the detection of radio emission from RXJ1856
would be a strong argument in favour of the model.

The effectiveness of the cyclotron mechanism has been estimated and
it appears to be quite efficient. For effective generation of waves
it is essential that the time during which the particles give energy
to waves should be more than $1/\Gamma$ (where $\Gamma$ is the
growth rate of instability). Generated radio waves propagate
practically in straight lines, whereas the dipolar magnetic field
lines deviate from its initial direction, and the angle
$\theta=k_{\parallel}/k_{\varphi}$ grows. $\theta$ is the angle
between the wave line and the line of dipole magnetic field. On the
other hand, the resonance condition (1) imposes limitations on
$\theta$ i.e. particles can resonate with the waves propagating in a
limited range of angles. The estimations show that the following
condition $\rho\gtrsim3\cdot10^{9}$cm should be fulfilled. As the
instability develops at the distances $r\sim10^{9}$cm, it follows
that the excited waves lie in the resonant region long enough for
particles to acquire pitch angles and generate the observed
radiation.

The total energy available for the conversion into pulsar emission
has been estimated, which approximately equals to $\dot{E} \simeq
n_{b_{0}} \pi R_{pc}^{2}\gamma_{b}mc^{3}$. Here, $n_{b_{0}}$ is the
Goldreich-Julian density at the pulsar surface and $R_{pc}$ is the
radius of the polar cap. The estimations show
$\dot{E}\simeq5\cdot10^{32}$erg/s, which is enough to explain the
observed X-ray luminosity of RXJ1856.

The recently discovered $7$s pulsations of the X-ray emission of
RXJ1856 has been explained in the framework of drift wave driven
model. The main feature of this model is that the spin period of
pulsar might differ from the observable period (for RXJ1856 the real
spin period is estimated to be $\sim1$s), as a consequence of the
existence of very low frequency drift waves in the region of
generation of the pulsar emission. These particular waves are not
detected but only result in a periodical change of curvature of the
magnetic field lines, which in turn cause the change of observed
radiation with a period of the drift wave.

In the present paper the emission polarization of RX J1856.5-3754 is
investigated, as we suppose that the measurement of the polarization
properties of RXJ1856 will make clear its real emission nature. It
was shown that both the X-ray and optical emissions are linearly
polarized. The degrees of polarization for the X-ray emission
approximately equals $\Pi_{\emph{l}}\approx8\%$ in the energy
interval of $(0.15 - 0.26)$keV and $\Pi_{\emph{l}}\approx(84-79)\%$
in the energy interval of $(0.26 - 0.9)$keV. The calculation of
degree of polarization for optical emission yields
$\Pi_{\emph{l}}\approx86\%$. If the emission of this source has a
thermal nature then according to \citet{ho077} (this model, among
other thermal emission models, gives the best match of the entire
spectrum) the polarization of X-ray emission should essentially be
equal to $100\%$. But if the emission of this source is generated by
the synchrotron mechanism, it is expected that both the X-ray and
the optical emissions will be linearly polarized with the frequency
dependent polarization degrees, giving the values from a few
percents up to $84\%$.

\section*{Acknowledgments}
The author is grateful to George Machabeli for valuable discussions.
This work was partially supported by Georgian NSF Grant ST06/4-096.
N.C. thanks the Abdus Salam International Center for Theoretical
Physics at Trieste, Italy.

\end{document}